\documentclass[a4paper,10pt]{article}
\usepackage[utf8]{inputenc}
\usepackage[T1]{fontenc}

\usepackage{graphicx}
\usepackage{wrapfig}
\usepackage{subfig}
\usepackage{hyperref}
\usepackage{units} 
\usepackage{authblk}
\usepackage{lineno}

\usepackage{xcolor}
\usepackage{xspace}

\newcommand{\tc}{Tumorcode\xspace}
\newcommand{\vbl}{VBL\xspace}
\newcommand{\po}{pO$_2$\xspace}
\newcommand{\pH}{pH\xspace}
\newcommand{\mum}{$\mu$m\xspace}

\title{Fine-grained simulations of the microenvironment of vascularized tumours}
\author[1,*]{Thierry Fredrich}
\author[1]{Heiko Rieger}
\author[2]{Roberto Chignola}
\author[3]{Edoardo Milotti}
\affil[1]{Center for Biophysics \& FB Theoretical Physics, Saarland University, Saarbrücken, 66123, Germany}
\affil[2]{Universita degli Studi di Verona, Department of Biotechnology, Verona, 37134, Italy}
\affil[3]{University of Trieste, Department of Physics,Trieste, 34127, Italy}

\affil[*]{thierry@lusi.uni-sb.de}

\begin{document}

\maketitle
\begin{abstract}
One of many important features of the tumour microenvironment is that 
it is a place of active Darwinian selection
where different tumour clones become adapted
to the variety of ecological niches
that  make up the microenvironment. 
These evolutionary processes turn the microenvironment into a 
powerful source of tumour heterogeneity and contribute to 
the development of drug resistance in cancer. 
Here, we describe a computational tool to study the ecology of 
the microenvironment and report results 
about the ecology of the tumour microenvironment and 
its evolutionary dynamics. 
\end{abstract}
%
%
\thispagestyle{empty}
\newpage


%
%
%
%
%
%
%
%
%
%
%
%

\section*{Introduction}

The tumour microenvironment is characterized by large chemical gradients and contains a mixture of normal and tumour cells. The presence of high gradients favours the formation of different ecological niches which can become an important source of tumour heterogeneity \cite{Anderson:2006aa,Sprouffske_2012,Beerenwinkel:2016aa,Yuan:2016aa,Basanta:2017aa,Gatenby:2018aa}. 

The actual size of the niches is quite small, as it is mostly determined by the structure of the scaffold of capillary vessels that envelope and feed the tumour mass. This structure is highly irregular, with a chaotic blood vessel network, a missing lymphatic network, elevated acidity, poor oxygenation, and high interstitial fluid pressure. The distance between blood vessels may be as large as a few hundred \mum, and the regions in between them can become highly hypoxic \cite{Yuan:2016aa}. 

The fine-graininess and the large spatial variability are essential features in the formation of the ecological niches and they must belong to any mathematical model that aims to explain tumour heterogeneity as the result of Darwinian selection driven by the local environment. 
The importance of the microenvironment is widely recognized, and several researchers 
have tackled the problem of its description and understanding from the computational
point of view,
e.g. 
\cite{
shirinifard_3d_2009,
tang_computational_2014,
grogan_microvessel_2017,
xu_mathematical_2016,
vilanova_mathematical_2017,
zhao_three-dimensional_2017,
zheng_nonlinear_2005}
,
and the reviews
\cite{
rejniak_systems_2016,
deisboeck_multiscale_2011}.

The studies, however, mainly focus on the biochemical, 
cellular and biophysical properties of the tumour microenvironment 
and not on its active role in providing varied evolutionary paths 
to genetic variants of the tumour cells. 
As it has already been pointed out 
\cite{gatenby_cancer_2011,gatenby_evolutionary_2014},
the adaptive evolution of tumour clones
(central concept of Darwinian dynamics)
is driven by the formation of new environmental niches.
Many practical difficulties limit the experimental 
study of the adaptation process,
while computer simulations can shed light --- albeit in a limited way --- 
on the dynamics of many steps like the convergent evolution of 
different genotypes to the same phenotype, and the selective loss of specific cell functions. 
\\
Simulations of avascular solid tumours show that 
the microenvironment of these small cell aggregates 
is formed by rather homogeneous niches with smooth gradients of oxygen, 
of other nutrients, waste molecules and cell viability \cite{milotti2010emergent}.
After the angiogenic transition, however, the microenvironment differentiates
in unpredictable ways. 
To identify underlying processes and provide reasoning,
we have developed a computational model
that is a tool to understand the fine-grained features of a simplified tumour microenvironment,
i.e. of a microenvironment that contains only tumour cells and blood vessels.
This is a rather strong simplifying assumption, 
since experimental observations show that in addition to 
tumour cells the microenvironment contains several 
other important cell types such as stromal and immune cells. 
We wish to stress that this assumption is a necessary compromise,
because no computer model can presently describe the huge biological 
complexity of the tumour microenvironment and capture its growth 
at appropriate spatial and temporal resolutions.

Our computational model combines two different simulation programs, 
namely, a lattice-free simulation of small avascular solid tumours 
and a lattice-based simulation of the blood vessels dynamics in solid tumours. 
As we discuss below, 
both models have been individually validated with experimental data, 
and here we provide a first validation of the combined model by 
correctly reproducing the relevant observed gradients of tumour oxygenation 
and other biochemical and biological features. 
Next, we use the combined model to explore how the tumour microenvironment 
takes shape in small solid tumours at the transition from the avascular 
to the vascular growth phases, 
the so-called angiogenic switch \cite{bergers_tumorigenesis_2003}. 
We follow the complex dynamics of this process in detail and in real time
and show that,
already at these early stages of tumour development, 
the diversity of environmental niches is high enough to promote adaptive evolution of tumour variants.


\section*{Brief outline of ``Virtual Biology Lab'' (\vbl) and \tc}

In this section we give a short description of the two programs that have been merged.
The present account is very brief as both programs have been described in detail in 
past papers (for \vbl, see
\cite{milotti2010emergent,chignola2012bridging,chignola2014single,milotti2012interplay,stella2014competing}, 
for \tc, see \cite{rieger_physics_2016,fredrich_tumorcode_2018,Welter_2016}). 

\subsection*{\vbl}
\begin{figure}
 \includegraphics[width=\textwidth]{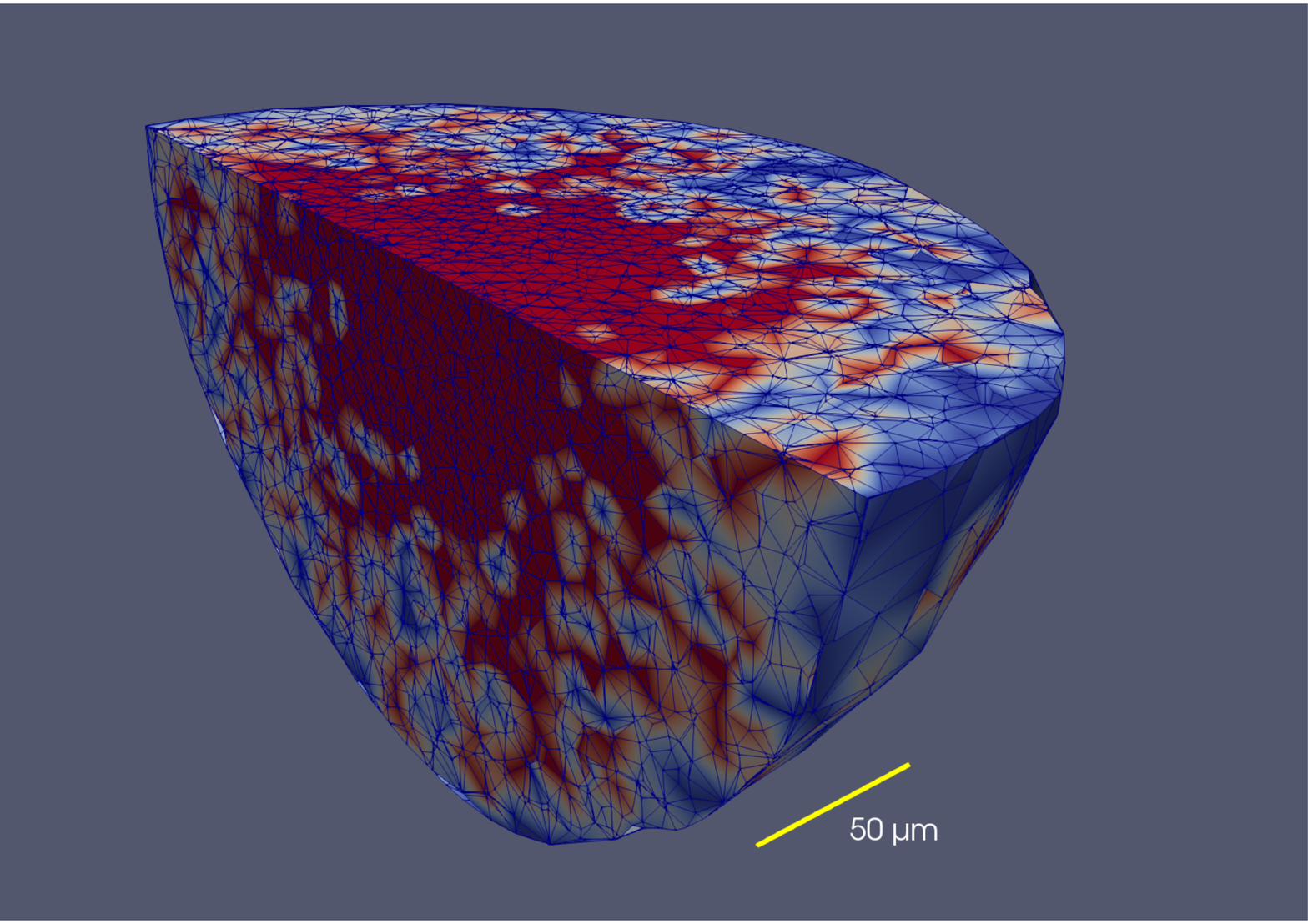}
 \caption{
 Distribution of cell phases in a tumour spheroid 
 (avascular \textit{in vitro} tumour) simulated with \vbl.
 The figure shows a clipped spheroid, to display the cell phases in the inside of the spheroid. The colors represent different cell phases, and in this example it is only important to note that the necrotic regions are red. Note that there is no clear cut boundary between the proliferating part of the necrotic core. The triangulation used by \vbl is also shown. This simulated tumour spheroid comprises almost 24000 cells.
 }
 \label{fig:phase_VBL_pie}
\end{figure}

\vbl is a lattice-free, cell-based program that can simulate both: 
the growth and proliferation of disperse cells and of more complex,
avascular cell clusters (tumour spheroids). 
It contains a detailed simulation of the metabolic activity of each cell
including discrete events at the individual cell level 
(for instance: mitochondrial partitioning at mitosis), 
and has the potential to activate phenotypic differentiation. 
More specifically, the model has the following features: 

\begin{itemize}
\item The model of human tumor cells includes both- the internal biochemical processes and a phenomenological description of the biomechanics of cells.
\item Spatially, each cell is a two-compartment structure, 
the inside of the cell and the adjacent extracellular environment
or intercellular space;
this is important to handle simple diffusion and facilitated diffusion across 
the cell membrane at the same time.
\item Discrete events are simulated and interleaved between successive deterministic steps; the random nature of some of them contributes to the stochasticity of the simulation as a whole.
\item Each cell is characterized by its own phenotype
which means that a specific set of parameters is linked to an individual cell.
\item Enzyme activity is modulated both by \po concentration and by \pH.
\item Weighting of molecular paths and cellular mechanisms is possible.
\item The program simulates dispersed tumor cells 
and cell clusters (tumor spheroids) including the intercellular space.
\item The surrounding environment is an integral part of the simulation. 
\end{itemize}

The numerical methods used in \vbl have been described elsewhere \cite{chignola2011computational},
here we only mention that although the simulation is lattice-free,
continuum processes are actually discretized on the continually-variable irregular 
lattice defined by the Delaunay triangulation \cite{de2000computational} 
based on the cells' centres which also defines the proximity relations among cells 
that are necessary to compute the cell-cell forces.
We find that the computational geometry library CGAL (\url{https://www.cgal.org}) \cite{cgal:eb-18b}
is an easy tool that offers necessary triangulation features
amended by the possibility to calculate alpha shapes. This is necessary 
to define the surface or contact zone of the spheroid with the environment.
It is important to note that the simulation spans many orders of magnitude in time
which makes the large system of differential equations 
describing the deterministic part of the program extremely stiff. 
Therefore we solve the equations by means of implicit integration methods, 
that allow for comparatively large time steps 
(of the order of 1-10 s in terms of simulated time) \cite{milotti2009numerical}. 

Figure \ref{fig:phase_VBL_pie} shows an example of a tumour spheroid 
simulated by \vbl. As we have shown in the past, 
the main metabolic, morphologic and kinetic parameters are correctly reproduced 
both for isolated cells and for cells grown as three- dimensional clusters
\cite{chignola2011computational,chignola2014single}.

\subsection*{\tc}
\begin{figure}
 \includegraphics[width=\textwidth]{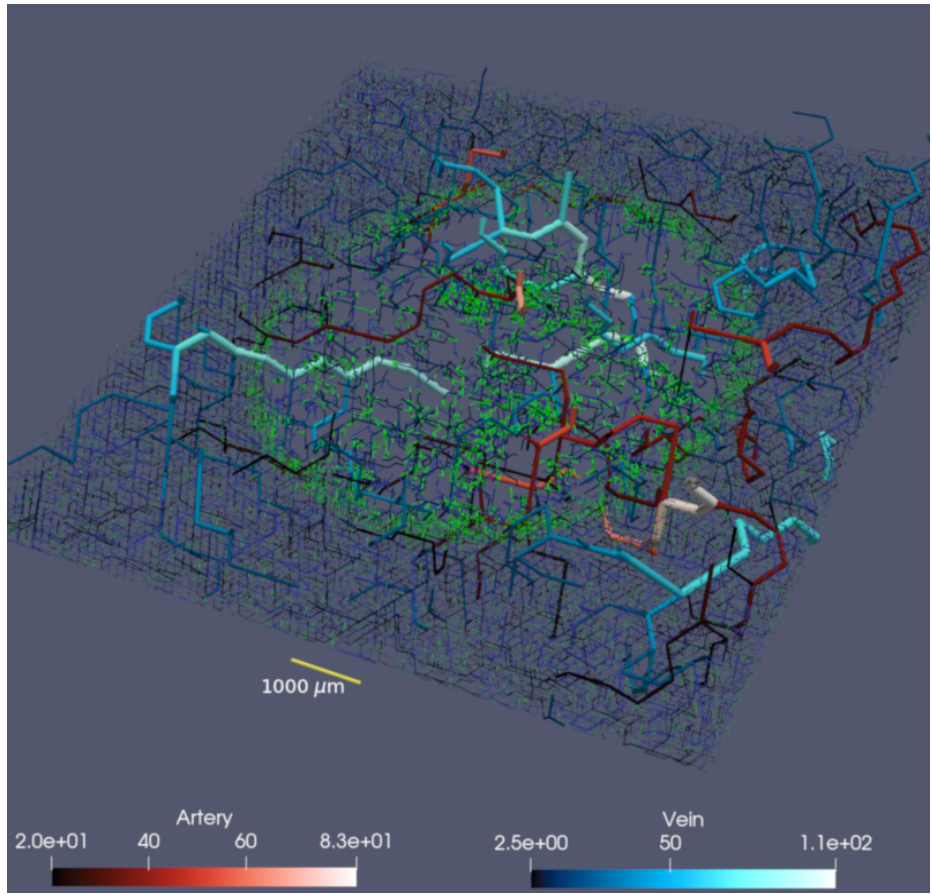}
 \caption{ 
 500 \mum thick slice trough tumour vasculature of 1cm lateral size. For 3D illustration 
 we extent vessels with radius bigger than 20 \mum for another 750 \mum above and below the slice.
 According to the lower panels: arteries are displayed in varying reddish color and 
 veins in varying blueish color (unit is also \mum). 
 Capillaries and young vessel are shown in green.
 }
 \label{fig:tumorcode_large}
\end{figure}

While cell-cell interactions like in \vbl naturally produce nearly spherical structures, 
modelling the growth of blood vessels moves the complexity of the simulation one step 
higher to the level of tissue morphogenesis. The list of models 
is long. For an appropriate discussion of the existing models
for tumor vasculature see \cite{Rieger:2015aa}.

Solid tumours grow in originally healthy tissue featuring a normal vasculature
build up by a hierarchically organized arterio-venous blood vessel network
that is then dynamically modified by the growing tumor.
An integrative modeling approach thus has to address two issues: 
first, it has to find an appropriate representation 
of the original vasculature of the host tissue;
and second, the dynamics of the given blood vessel network has to be defined,
which includes the insertion of new vessels via angiogenesis 
as well as the removal of existing vessels (old and new) 
via vessel regression, and the modification of existing vessels via dilation, 
constriction or occlusion. 
Moreover, the network carries a blood flow and transports and emits 
oxygen, nutrients and drugs, which has to be represented, too. 
Various modeling approaches have been presented in the past 
(
for a review see e.g. 
\cite{Rieger:2015aa,Welter:2016aa,rieger_physics_2016}
), here we use the simulation framework \tc
described in \cite{fredrich_tumorcode_2018}, 
which does not only create artificial arterio-venous 
blood vessel networks representing the healthy initial vasculature, 
but includes vascular remodeling during solid tumour growth by angiogenesis, 
vessel dilation, regression and collapse. 
The created networks were tested against experimental data 
and successfully reproduced: 
1) morphology and flow characteristics \cite{welter_vascular_2009,welter_physical_2010},
2) interstitial fluid flow and pressure \cite{Welter:2013aa}, and 
3) tissue oxygenation \cite{Welter_2016}
for both, healthy and tumour tissue.

Again, here is a short list of its main features of \tc:

\begin{itemize}
\item Hemodynamics includes the phase separation effect (F{\aa}hr{\ae}us-Lindqvist effect)
and different rheologies.
\item Concentrations in the surrounding environment are computed according to a continuum model,
i.e., according to a set of partial differential equations discretized 
by finite elements method solved with the 
Trilinos C++ library (\url{https://trilinos.org}) \cite{Heroux_2005}.
\item Angiogenesis, i.e., the addition of new segments, is driven by the VGEF gradient and is partly stochastic \cite{rieger_physics_2016}.
\item As the tumour grows, the vasculature and the blood flows change,
and the modified shear stress leads to \textit{blood vessel dilation}
or to \textit{blood vessel collapse}
\cite{rieger_physics_2016}. 
\item A low local VGEF growth factor concentration produces \textit{blood vessel regression} \cite{rieger_physics_2016}.
\item The program computes the interaction with surrounding tissues to compute the extraction of oxygen and nutrients from blood vessels.
\item The program returns blood flow, oxygen concentration, metabolite concentration, etc, in blood vessels and surrounding tissues.
\item The blood vessel hematocrit acts as source for oxygen dissolved in tissue and the tissue  simultaneously consumes the oxygen. In \cite{welter_computational_2016} we explain how the coupled set of non linear differential equations is solved to obtain the oxygen partial pressure for each blood vessel segment.
\end{itemize}

Figure \ref{fig:tumorcode_large} shows an example of a tumour vasculature simulated by \tc. 
You can clearly distinguish the vasculature inside the tumour from the surrounding healthy tissue. 
Note the difference in
length scale when comparing Figure \ref{fig:tumorcode_large} and Figure \ref{fig:phase_VBL_pie},
and the fact that the tumour is a continuous structure 
( governed by reaction-diffusion equations \cite{fredrich_tumorcode_2018})
in \tc. In contrast to \vbl, it cannot describe the evolutionary features 
of the microenvironment at the single cell level.

%
%
%

\section*{The merging of the two programs}

\begin{wrapfigure}{r}{0.5\textwidth}
\centering
\includegraphics[width=0.5\textwidth]{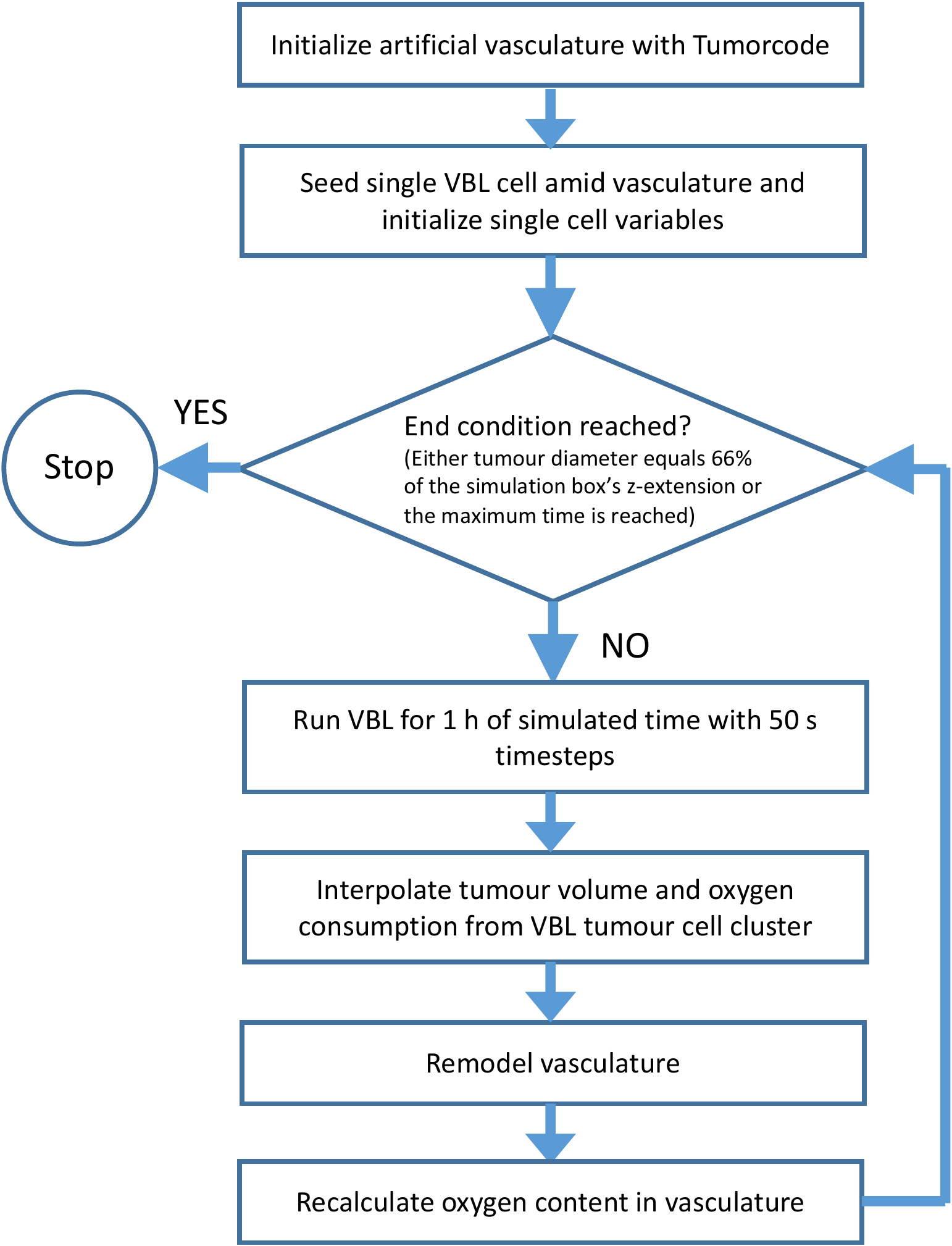}
\caption{Flowchart of combined \vbl and \tc simulation}
\label{fig:combined_flow}
\end{wrapfigure}

To merge the two programs into one, we had to match the spatial structures
of the two programs. The cells are not confined to a lattice however 
the diffusion processes in \vbl are calculated by means of a disordered 
Delaunay triangulation which is  not permanent 
and has a variable size, as cells position change at each time step while 
the cells grow and proliferate. The situation is almost the reverse in \tc
where blood vessel segments belong to an FCC lattice and the continous space 
is discretized by a cubic lattice to solve the partial differential 
equations. We interpolated cell properties from the cells position to 
the surrounding nodes of the cubic lattice (compare Figure \ref{fig:combined_flow}).
In particular, the \tc
requires an oxygen consumption field and each cell is considered as point- like source 
of growth factor. Once the growth factor field exceeds
the threshold, the vessel remodelling of \tc starts.



\medskip 

We estimate that the accurate calculation of any substance concentration
for a given cell requires 
$\mathcal{O}(N_\mathrm{C} N_\mathrm{BV})$
number of operations, 
where $N_\mathrm{C}$ is the number of cells and 
$N_\mathrm{BV}$ is the number of blood vessels segments. 
Since both $N_\mathrm{C}$ and $N_\mathrm{BV}$ easily exceed $10^6$ 
in the smallest-sized biologically relevant simulation runs, 
$\mathcal{O}(N_\mathrm{C} N_\mathrm{BV}) \sim 10^{12}$ per time step, 
making the simulations utterly unmanageable. 
To reduce the computational complexity, we 
take into account that only the nearest blood vessels actually
contribute to the inputs and outputs of the different molecules
in the vicinity of a given cell trimming the large number of 
cell-blood vessel pairs to $\mathcal{O}(N_\mathrm{C})$.
Moreover, the timesteps to follow cell biochemistry are of 
order 1-10 s while biological relevant timesteps for vascular remodelling 
are in the order of hours. Therefore we introduce two different time 
steps to avoid a huge number of useless sweeps over the blood vessel network
(compare Figure \ref{fig:combined_flow}).


%




\section*{Development of the microenvironment in small solid tumours}

\begin{figure}
 \centering
 \includegraphics[width=0.7\textwidth]{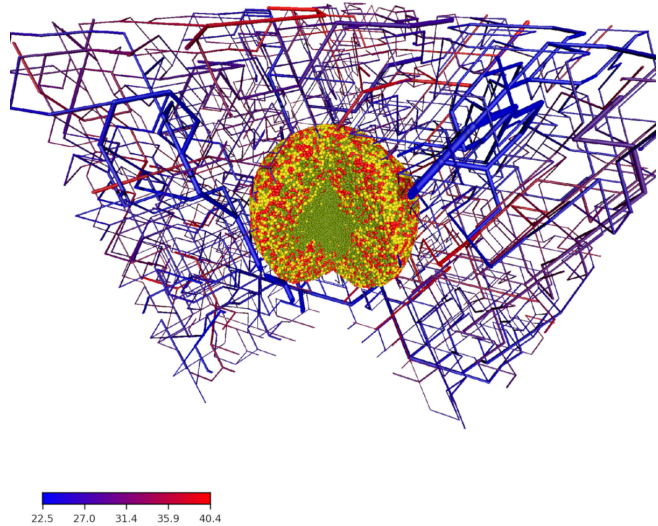}
 \caption{
 Artificial blood vessel network containing a 
 tumour after 23 simulated days past the initial seeding. 
 For visualization we removed a quarter of the system. 
 The color code in the  bottom left corresponds to the blood pressure at a given vessel segment. Cells are colored by their cell phase. Red: phases G1m and G1p; yellow: phases S, G2 and M; light green: dead cells
 }
 \label{fig:cell_phases_pie}
\end{figure}

With the program described before, we studied selected aspects of the tumour microenvironment.
To this end, we carried out simulation runs with three different tumour seeds: 
1. at the center of a blood vessel network, 
2. close to an arterial bifurcation,
3. close to a venous bifurcation. 
The details of each simulation are reported in the supplemental material.
Here we display the results in a way that allows
an easy comparison with measurements \cite{Yano:2014aa,helmlinger_interstitial_1997}.


As expected, the seeding is important in the very first steps of tumour growth: 
a seed close to blood vessels leads to fast growth and to a quick remodelling 
of the blood vessel network;
for seeds far from blood vessels, the initial growth is
slower and the subsequent initial development of the tumor microenvironment differs slightly. 

\subsection*{Seed at the center of a blood vessel network}

\begin{figure}
 \centering
 \includegraphics[width=0.5\textwidth]{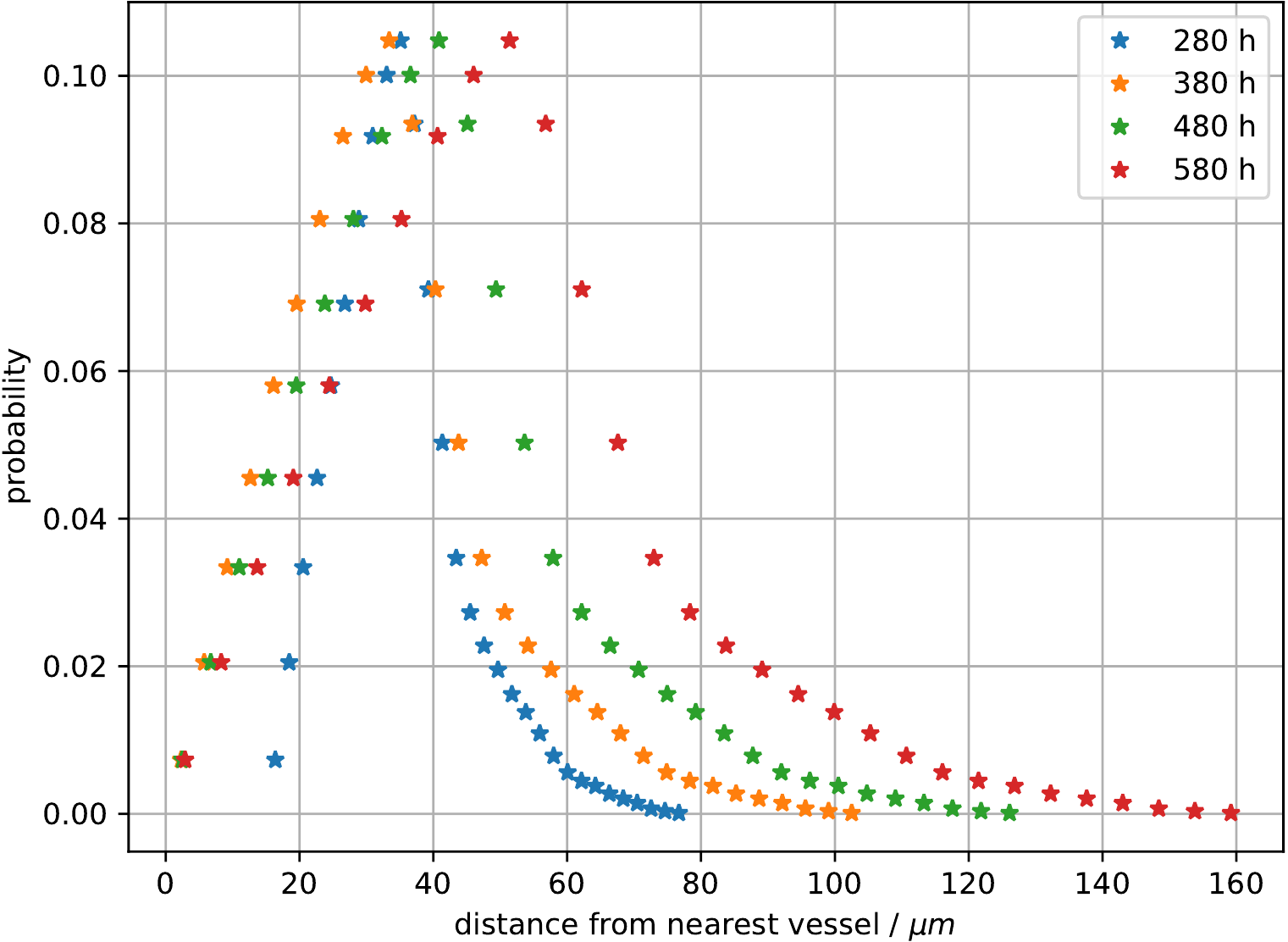}
 \caption{
 Empirical distribution of the distance to the closest vessel for all cells in the developing tumour.  The snapshots are taken at simulated times 280 hours, 380 hours, 480 hours and 580 hours past the initial seed. Colors represent the different simulation times. 
 }
 \label{fig:prop_dist_multiple}
\end{figure}

The computational approach provides an easy access to information that is 
hard to obtain with experiments and allows multiple ways of visualization 
(compare Figure \ref{fig:cell_phases_pie}). 
Here we take advantage of the information about the distance to the nearest blood vessel 
for every cell.
We start with the distribution which is a defining feature of the microenvironment,
as larger distances mean less oxygen and nutrients, 
and therefore starvation and death for many tumour cells. 
In Figure \ref{fig:prop_dist_multiple} we see how the distribution changes while the 
tumour grows.
The distribution at the early stage ( 280 hours past seeding) ranges from 
0 to 80 \mum ensuring a optimal oxygenation to all cells and gets distorted 
until it covers a range of up to 160 \mum at a later stage ( 580 hours past seeding)
where cells no receive a sufficient input of oxygen and nutrients.



\begin{figure}
\centering
\includegraphics[width=0.5\textwidth]{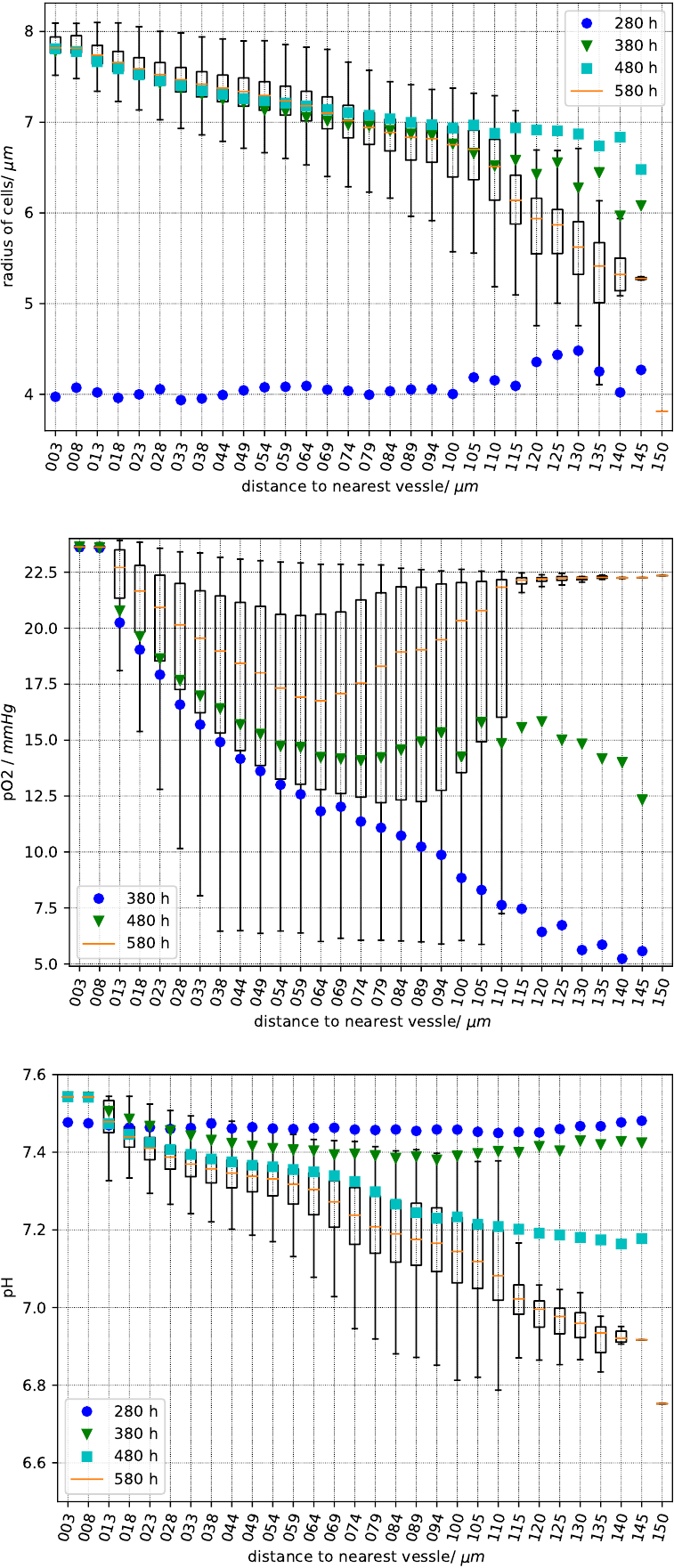}
\caption{  
Sampled cell quantities vs. the distance to the nearest vessel.
Top panel: cell radius, center panel: partial oxygen pressure,
bottom panel: extracellular \pH.
Each box plot has been obtained using the values in cells at a given distance 
from the nearest vessel as indicated by the abscissa. 
As usual, the vertical error bars (``whiskers'') represent the lower and upper quartile.
The data is sampled 280, 380, 480 and 580 hours past the initial seeding.
For better clarity, we omit the complete boxes for 
the time points 280, 380 and 480 
and plot only the mean value in different colors.
}
\label{fig:sampled_distance_to_nearest}
\end{figure}

Intuitively the increase in the mean of cell-blood vessel distances leads
to a decrease of available oxygen and nutrients, 
and to an increase of lactate ions and to a lower \pH. 
Since the cell-blood vessel distance is known, we can
display space-time data on the distributions of cell radii, \po and \pH
(Figure \ref{fig:sampled_distance_to_nearest})
- the data confirms the intuition. 
We uploaded two videos where we visualize 
\po 
(\verb|fredrich:po2:2018|)
and 
\pH 
(\verb|fredrich:pH:2018|)
during the the complete of the growth processes.



\begin{figure}
 \centering
\includegraphics[width=0.8\textwidth]{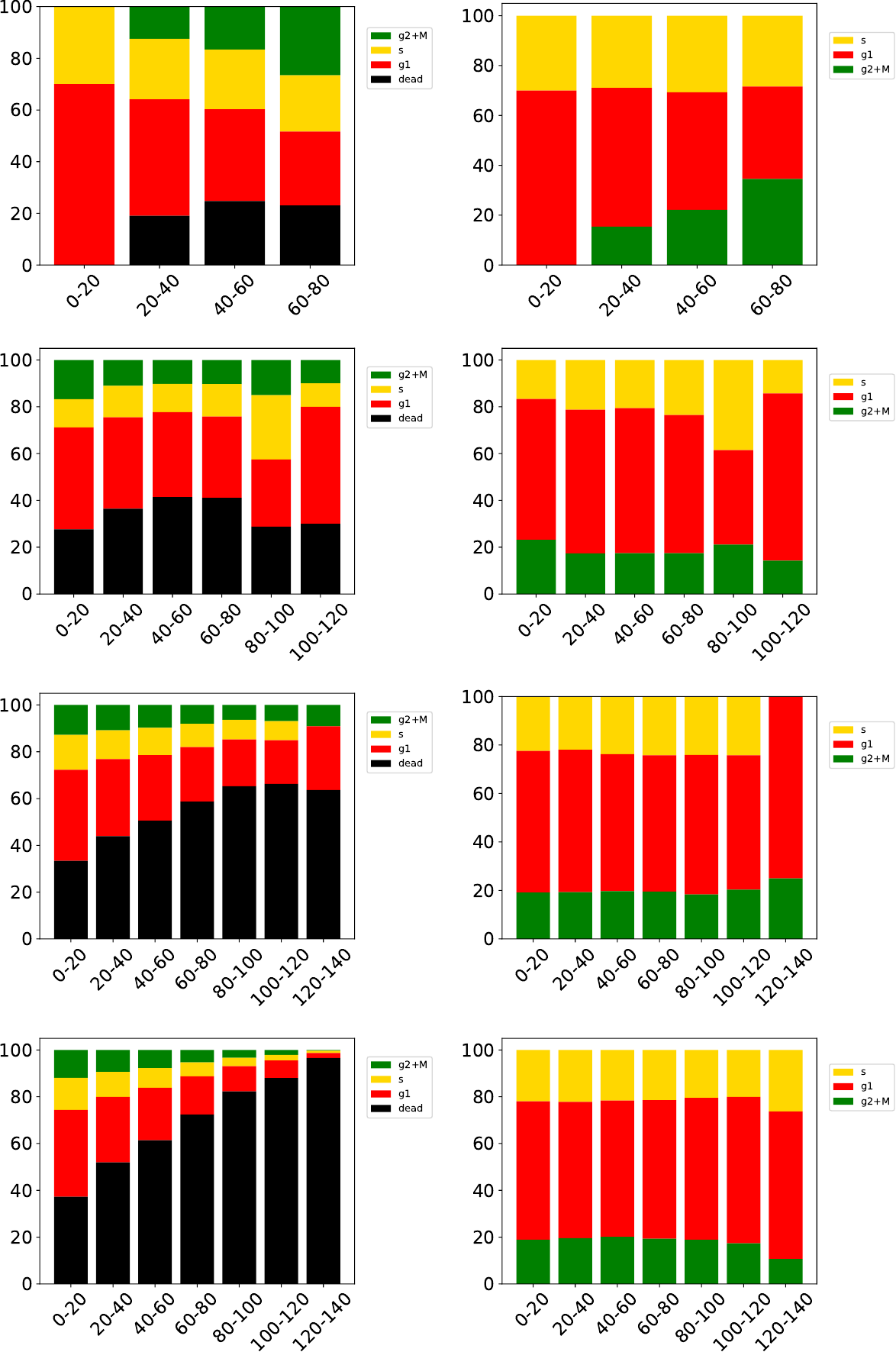}
\caption{  
Distribution of the cell phases at different distances from the closest blood vessel.
From top row to bottom row the data is take after 280, 380, 480, 580 hours past
this inital seed. The left panels show the number of cells in the corresponding phase
relative to the total number of current cells. G1m and G1p in red, S in yellow, 
G2 and M in green and dead cells in black. In the right panels we consider 
only alive cells.
The abscissa is the distance to the nearest vessel.
}
\label{fig:cell_phases_from_nearest_vessel}
\end{figure}

Finally, the changes in the microenvironment stretch the cell proliferation period
and therefore modifies the cell cycle distribution as seen in Figure 
\ref{fig:cell_phases_from_nearest_vessel} where we can follow the formation
of a necrotic core. At the beginning (280 hours past initial seed) the 
relative amount of cells in the different phases varies with the distance 
from the nearest vessel. Surprisingly this relative amount stays almost 
constant after the establishment of different niches.
Compare with Figure \ref{fig:cell_phases_pie} were we show the cell phase 
distribution in a more pictorial way. Even in such a small tumour we observe 
the initial development of necrotic regions which characterizes the development of
the tumour microenvironment.



%

Actual tumours have irregular shapes, 
while the cells in \vbl are not polarized, 
and when grown in uniform environments produce nearly spherical shapes.
In the presented case the blood vessel network drives the growth
in specific directions and we expect a marked deviation from 
sphericity already with small simulated tumours.
Ideally, in a spherical tumour all cells on the surface have the same distance 
to the centroid. We find that the distribution of distances from the spheroid's surface
to the centroid a) broadens and b) deviates from symmetric form as the tumour grows
which is a clear indication of the deformation of the cell cluster and of its deviation from sphericity. 

\begin{figure}
 \centering
 \includegraphics[width=0.7\textwidth]{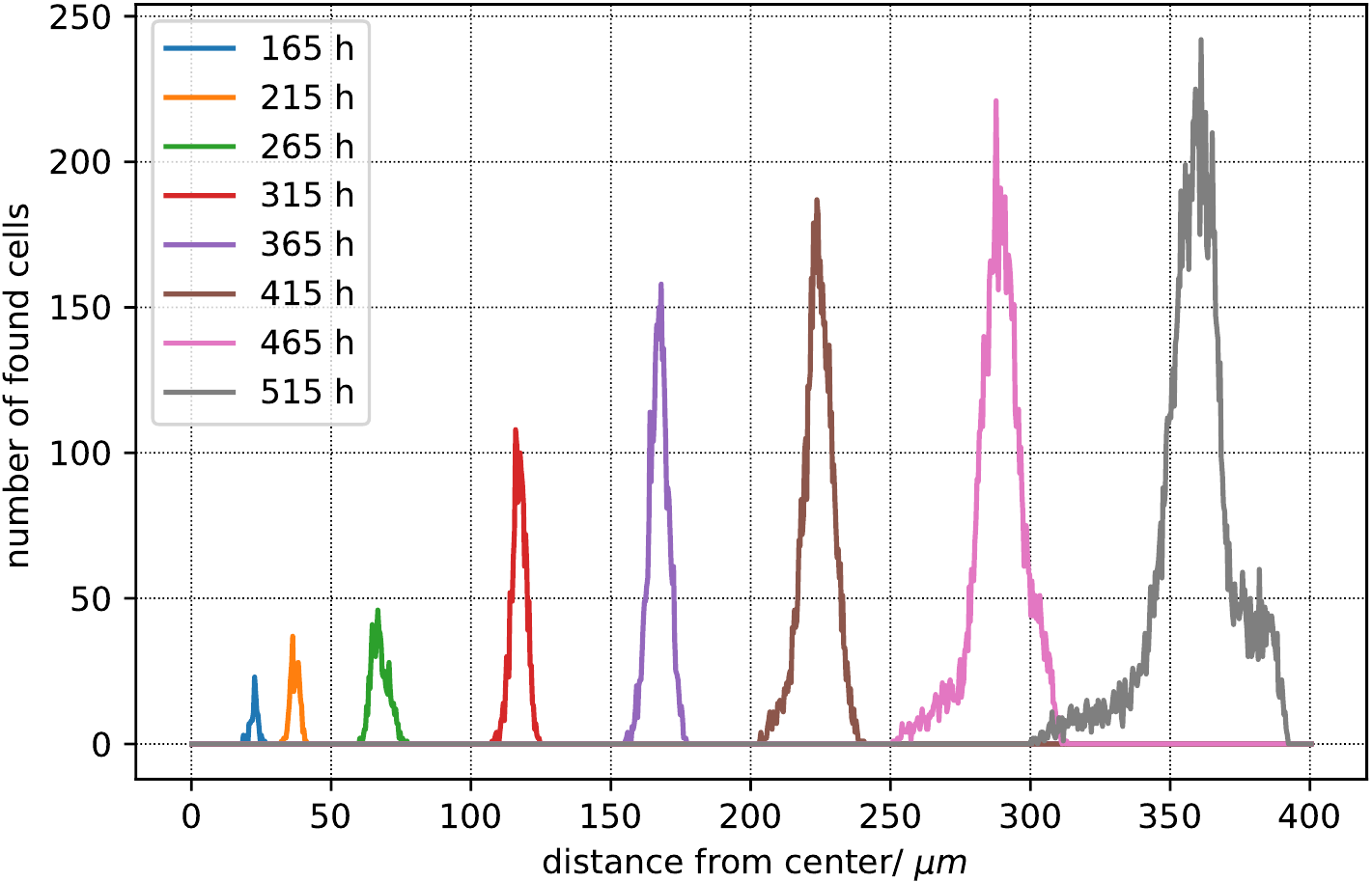}
 \caption{Distribution of distance from tumour centroid 
 to cells in contact with the environment (normal tissue). 
 The colors indicate the different sampling times past the inital seeding. 
 The bin size is 30 \mum.
 }
 \label{fig:dev_from_sphere}
\end{figure}


\subsection*{Seed close to an arterial or venous bifurcation}

\begin{figure}
 \centering
\includegraphics[width=\textwidth]{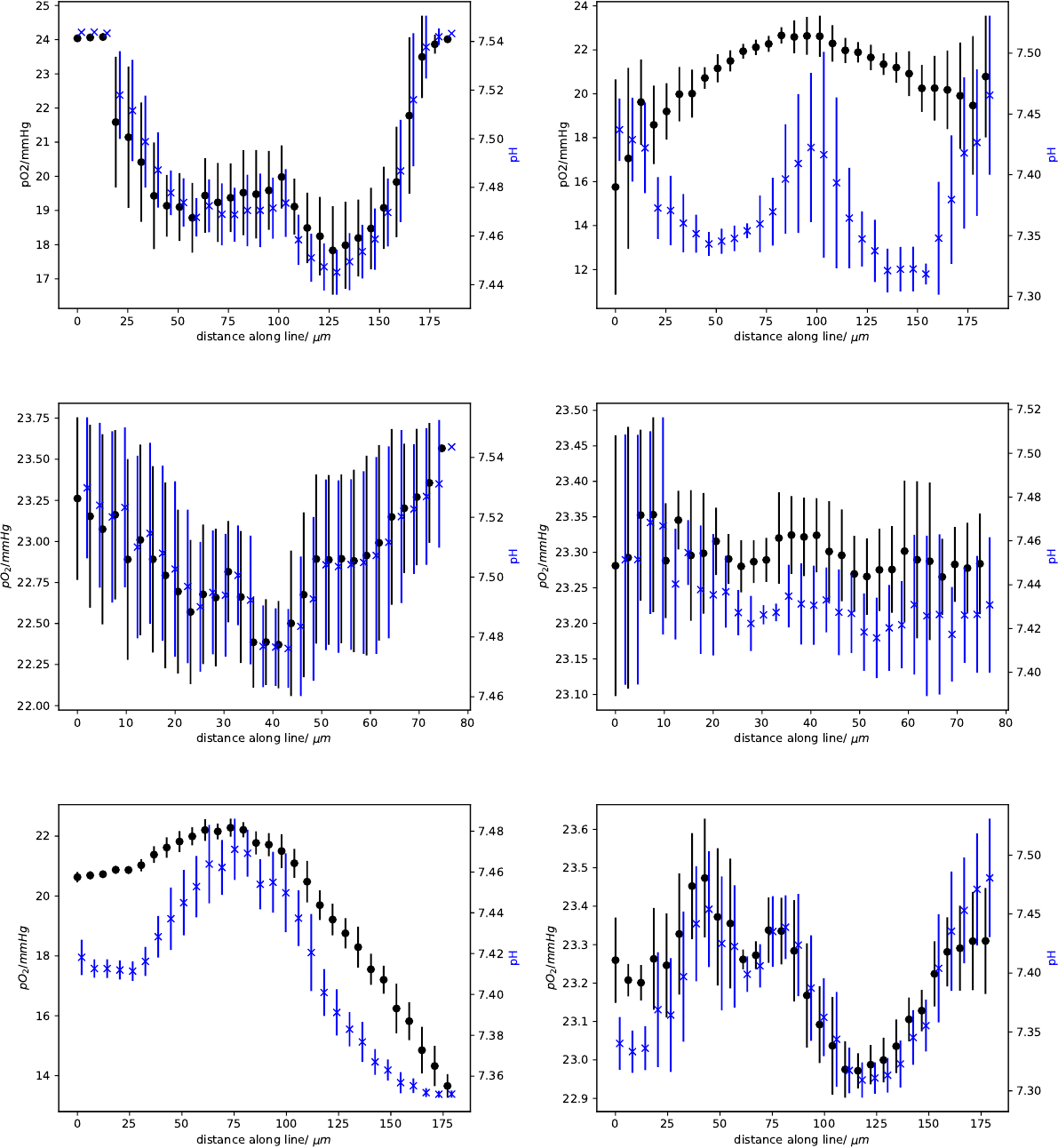}
  \caption{
  Data for the three measured lines. Each row corresponds 
  to one specific line.
  The left column represents data taken at 350 hours of simulated time, and the 
  right column at 495 hours past the initial tumour seeding.
  Each panel shows the sampled partial oxygen pressure 
  (\po) in 
  black (left hand y-axis) and the \pH in blue 
  (right hand y-axis).  
  To improve visibility, the data points of \po and \pH 
  are shifted against each other by 2 \mum. 
  Each line was sampled by 30 points within a sphere of radius 10 \mum. 
  The vertical lines represent the standard deviation.  
  }
  \label{fig:all_cases}
\end{figure}

Helminger et al. \cite{helmlinger_interstitial_1997}
measured the partial oxygen pressure and \pH within 
different tumours.
To compare our simulations with their experimental results,
we seeded the initial tumour cell close to bifurcations,
both arterial and venous, and sampled the cells
along different lines (for details see supplemental material).
Figure \ref{fig:line_at_350_and_527} is meant to 
illustrates the sampling procedure
while we show data for a early and a late simulation state
in Figure \ref{fig:all_cases}.

\begin{figure}
 \centering
 \includegraphics[width=\textwidth]{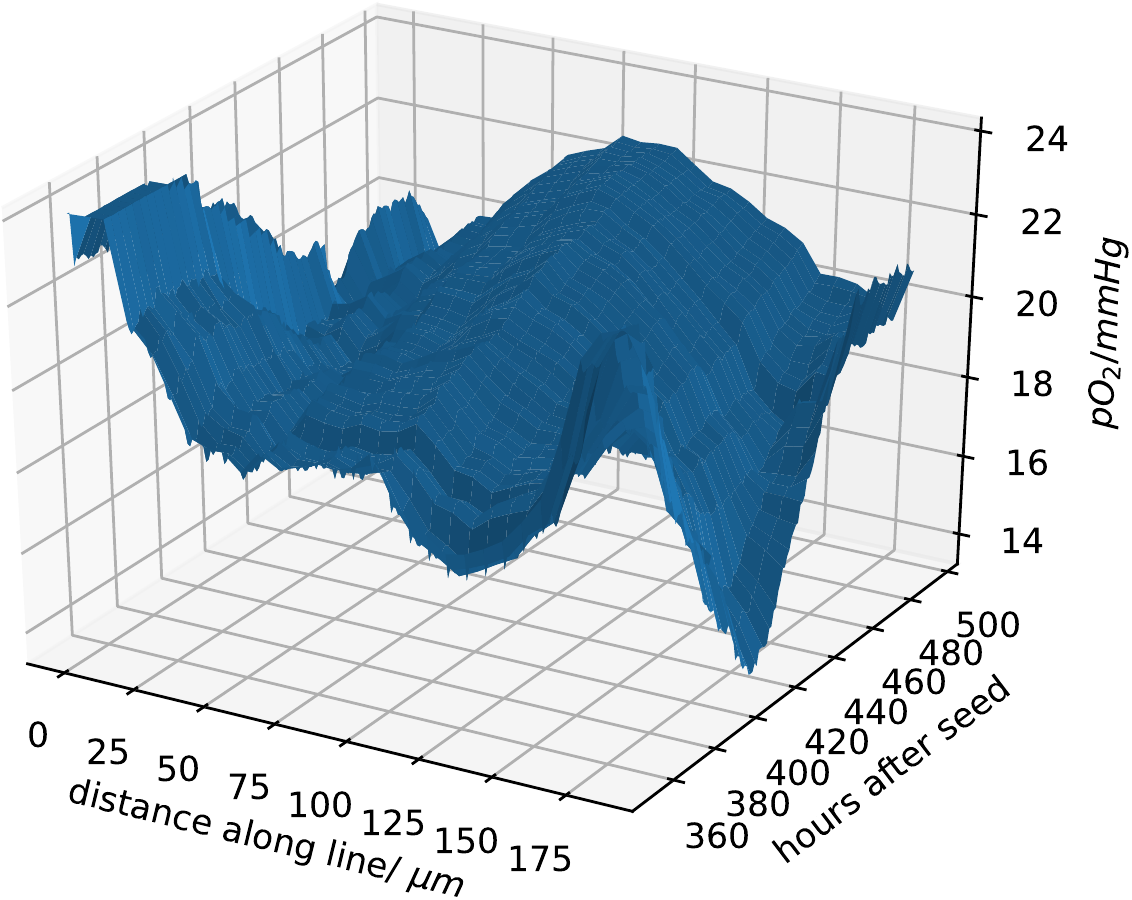}
  \caption{Change of \po  due to the blood vessel dynamics. This data extends those in the first row of Figure \ref{fig:all_cases}.
  In contrast to Figure \ref{fig:all_cases}, we show all 
  intermediate time points in this plot.
  }
 \label{fig:po2_dynamics}
\end{figure}

It is interesting to note that the left panel in first row 
in Figure \ref{fig:all_cases} shows a dip in both 
\po and \pH approximately midway between 
the two blood vessels. 
The right panel in the same row, taken at a later time,
shows a maximum at roughly the same position
because a new blood vessel sprouted into the intervening space.
A detailed view of this dynamic change is shown in 
Figure \ref{fig:po2_dynamics}, where the mean of the \po data 
is plotted for all intermediate time points as the new 
blood vessel grows.


\begin{figure}
 \centering
 \includegraphics[width=0.7\textwidth]{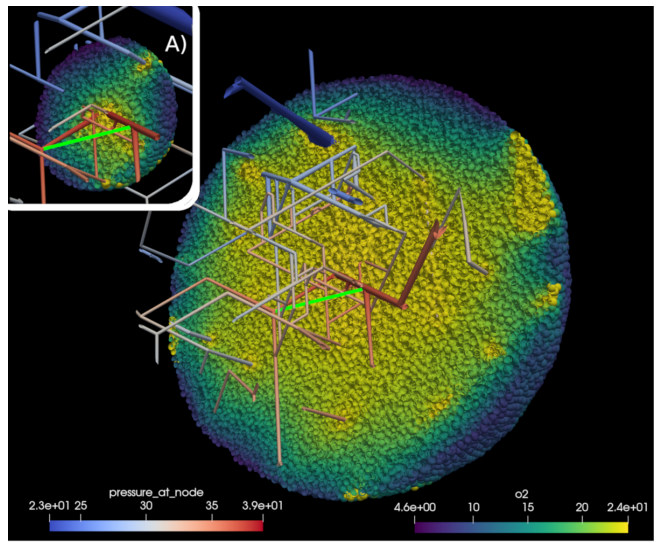}
 \caption{ 
 Sampling the partial oxygen pressure \po along lines between blood vessels.
 We sampled along the line joining two blood vessels (bright green). 
 Inset A) 
 shows the simulation state 350 hours past initial seeding;
 the main image shows the state after 527 hours. 
 Note the additional vessel due to angiogenessis. 
 The scale marked ``pressure\_at\_node'' refers to the \po value at vessel nodes in mmHg. The scale marked ``o2'' referes to the value of \po inside cells.
 See also supplemental material for more information.
 }
 \label{fig:line_at_350_and_527}
\end{figure}

\section*{Conclusions}
As explained in the Introduction, 
the study of the ecology of the tumour microenvironment 
at the angiogenic switch is 
the primary motivation for the work described in this paper 
and for the development of the presented complex software tool. 
Indeed, the Figures \ref{fig:all_cases} and \ref{fig:po2_dynamics}
display a surprisingly large spatial 
and temporal variability already in a very small vascularized tumour. 
In particular, Figure \ref{fig:po2_dynamics} shows a continually changing
and rugged landscape:
this means that the niche diversity is large, 
and consequently that there is a high evolutionary pressure, 
and a high probability that the microenvironment selects 
different tumour clones even in small tumours. 
This is important in view of the long time required for tumour growth,
as it has been estimated that a tumour requires about 10 years 
to reach a size of 1 cm \cite{talmadge2007clonal}.
According to the \textit{clonal selection hypothesis} 
new and aggressive tumour clones develop through 
Darwinian selection during this extended growth time 
\cite{talmadge2007clonal,arneth2018comparison}.
The results discussed in this paper lend further credibility 
to the \textit{clonal selection hypothesis}, 
and in our future studies we plan to demonstrate 
the spatial and temporal variations of different clone populations 
in this very complex tumour microenvironment.
In order to highlight common and 
diverse evolutionary pathways in different cancers, 
we also plan to explore the adaptive evolution of cell clones 
in different microenvironments mimicking different 
organs or tissue types
to reproduce 
the distinguishing features of selected tumour types.

Finally, we note that the molecular mechanisms 
that promote genotypic changes in tumour clones 
are well known and understood, 
but genotypic variability is only one feature of cancer's evolutionary landscape.
The other important feature is the variability of the environment 
that supervenes the genome and drives evolution itself, 
and with our computational tool we can start to explore its 
Darwinian dynamics \cite{gatenby_cancer_2011}.



\bibliography{bib}
\bibliographystyle{plain}


%
%
%
\section*{Acknowledgments}

The authors thank the Collaborative Research Center SFB 1027, 
the University of Trieste (FRA 2016), 
the University of Verona (RATs 2015), 
and the Deutscher Akademischer Austauschdienst (DAAD), 
for providing financial support to this project.  
Finally, we thank Benjamin Bogner for helping us with the computer graphics. 

\section*{Author contributions statement}
TF: conducted simulations, analyzed data, drafted results part of manuscript;
EM: suggested scientific problem, added VBL to public domain, drafted other sections;
TF and EM: designed an interface for VBL and Tumorcode, optimized parameters;
RC: helped with biological aspects and parameter tuning;
HR: developed Tumorcode program, supervised TF. 
All authors approve the content and reviewed the manuscript. 

\section*{Additional information}
\textbf{Supplementary information:}\\
Simulation details are described in a supplementary file.\\
\textbf{Source code availability:}\\
All source code containing the used parameters is distributed with this article
as supplementary material
and hosted at GitHub
\href{https://github.com/thierry3000/tumorcode}
{https://github.com/thierry3000/tumorcode}. \\
\textbf{Data availability:}\\
We uploaded the full simulation raw data to zenodo with 
following digital object identifiers:\\
\href{https://doi.org/10.5281/zenodo.2541688}
{https://doi.org/10.5281/zenodo.2541688}\\
\href{https://doi.org/10.5281/zenodo.2541655}
{https://doi.org/10.5281/zenodo.2541655}\\
\href{https://doi.org/10.5281/zenodo.2541678}
{https://doi.org/10.5281/zenodo.2541678}\\
\href{https://doi.org/10.5281/zenodo.2541667}
{https://doi.org/10.5281/zenodo.2541667}
\\
\textbf{Competing interests:}\\
The authors declare no competing interests.

\end{document}